\documentclass[sigconf,nonacm]{acmart}

\usepackage[utf8]{inputenc}
\usepackage{paralist}
\usepackage{subcaption}
\usepackage{tabularx,array,booktabs}

\title{Reliabuild: Searching for High-Fidelity Builds Using Active Learning}

\author{Harshitha Menon$^*$, Konstantinos Parasyris$^*$, Tom Scogland, Todd Gamblin}
\email{{harshitha,parasyris1, scogland1, tgamblin}@llnl.gov}
\affiliation{%
  \institution{Lawrence Livermore National Laboratory}
  \streetaddress{P.O. Box 808}
  \city{Livermore}
  \state{CA}
  \country{USA}
  \postcode{94550-0808}
}

\usepackage{natbib}
\usepackage{graphicx}
\usepackage{bm}
\usepackage{algorithm,algpseudocode}
\usepackage{cleveref}
\usepackage{xcolor}
\usepackage{hyperref}

\hypersetup{
colorlinks=true,
linkcolor=blue,
urlcolor=blue
}

\definecolor{MyBlue}{rgb}{1,0,0}

\newcommand{\argmax}{\operatorname*{arg\,max}}

\newcommand\EI{\mathcal{I}} 
\newcommand\PG{p_g} 
\newcommand\PB{p_b} 
\newcommand\HH{\mathcal{H}}  
\newcommand\MM{\mathcal{M}} 
\newcommand\xx{\bm{x}}

\newcommand\tool{\emph{Reliabuild}}

\begin{document}
\begin{abstract}
Modern software is incredibly complex.
A typical application may comprise hundreds or thousands of reusable components.
Automated package managers can help to maintain a consistent set of dependency versions, but ultimately the solvers in these systems rely on constraints generated by humans.
At scale, small errors add up, and it becomes increasingly difficult to find high-fidelity configurations.
We cannot test all configurations, because the space is combinatorial, so exhaustive exploration is infeasible. 

In this paper,
we present \tool{}, an auto-tuning framework that efficiently explores the build configuration space and learns which package versions are likely to result in a successful configuration.
We implement two models in \tool{} to rank the different configurations and use adaptive sampling to select good configurations
with fewer samples.
We demonstrate \tool{}'s effectiveness by evaluating 31,186 build configurations of 61 packages from the Extreme-scale Scientific Software Stack~(E4S). \tool{} selects good configurations efficiently.
For example, \tool{} selects $3\times$ the number of good configurations in comparison to random sampling for several packages including Abyss, Bolt, libnrm, OpenMPI.
Our framework is also able to select all the high-fidelity builds in half the number of samples required by random sampling for packages such as Chai, OpenMPI, py-petsc4py, and slepc.
We further use the model to 
learn statistics about the compatibility of different packages, which will enable package solvers to better select high-fidelity build configurations automatically.
\end{abstract}

\maketitle

\def\thefootnote{*}\footnotetext{These authors contributed equally to this work.}\def\thefootnote{\arabic{footnote}}
\section{Introduction}

Since at least the 1960's, developers
have striven to make use of software components~\cite{mcilroy1968mass}.
Reusing components saves time and separates concerns---client code can
rely on robust implementations of common functionality without reimplementing them. 
However, with the efficiency of reuse
has come an increase in complexity~\cite{lehman1980programs, gonzalez2009macro, decan2019empirical,bogart2016break, dietrich2019dependency}. 
Package management is a cornerstone of modern software engineering; millions of components, or {\it packages} are available from public registries and can be included in a project by simply running a command or modifying a line in a file.

To deal with this complexity, software ecosystems today use automated package managers (e.g., APT, NPM, Maven, Cargo, and Spack), which analyze compatibility constraints among packages and select a consistent set of versions to install.
Simply selecting a {\it compatible} 
version configuration is known to be NP-complete~\cite{dicosmo:edos,mancinelli+:ase06-foss-distros}, but there may be many valid configurations.
Selecting the {\it best} of these requires that we solve an NP-hard constraint problem {\it along with} 
an NP-hard optimization problem~\cite{michel+:lococo2010,argelich+:lococo2010,tucker:opium}. 

There are many reasons to choose software versions
carefully.  Recent exploits in package ecosystems have highlighted issues with the strategies employed by most package managers~\cite{dependency_confusion}~\cite{npm_colors}.  In particular, many systems will choose the
latest possible version, assuming that it has the latest
fixes and security updates. However, using the latest version can also break package builds. We would like to be able to balance these concerns, but threading the needle between the need to track updates and choosing known stable builds is extremely hard. The alternatives are to use \emph{wisdom of the crowds}
(choice of the majority)~\cite{mileva2009mining} when selecting package versions or to select the \emph{lowest} allowed version.
However, none of these prevent or detect  conflict defects~\cite{artho2012software}.

In the scientific software and ML ecosystems,
the pain of build
errors is particularly acute. Code is written in
in compiled languages such as C, C++, and Fortran, along with interpreted front-ends in Python and Lua.
There are many compilers, package versions, language versions, and interoperability concerns.
Porting to new systems can lead to a myriad of errors.
We would like our packaging tools to help by visiting the bleeding edge to find bad configurations before we encounter them. 
Ideally, we could learn from related builds and increase the likelihood that a build we have never tried before will work.
We call such builds ``high-fidelity builds``.


In this paper, we present an active-learning-based method 
to select configurations with high-fidelity, and thus a high likelihood of building successfully. Our approach uses adaptive sampling, a technique where samples are
collected in an iterative fashion, to reduce the number of samples used to identify high-fidelity built configurations.
We leverage the inherent relationship between sub-packages in the form of a dependency graph and use it to design a 
surrogate model to predict whether a configuration will successfully compile. Our surrogate model is a probabilistic model
that gives a joint distribution over the dependencies to calculate a score that indicates whether a configuration will
build. This provides a set of promising samples which are used by the adaptive sampling algorithm to evaluate their 
true objective function by running the build process. This approach will enable users to select configurations that
are highly likely to build using limited evaluations, reducing the user effort and resource overhead.
Our main contributions are:
\begin{itemize}
    \item \tool{}, a novel active-learning-based approach that selects high-fidelity build configurations.
    \item Two probabilistic surrogate models that assign builds scores using: 1) wisdom of the crowd, 2) a Bayesian model that learns from dependencies among the packages.
    \item Automatically deriving dependency version constraints by analyzing the relative importance of different packages in the learned model.
    \item A detailed experimental evaluation of our approach over 31,186 builds from the E4S scientific software ecosystem.
\end{itemize}

In our evaluation, we find that \tool{} selects high-fidelity build configurations
automatically using far fewer samples than a randomized selection. The Bayesian as well as the wisdom of the crowd
surrogate models are able to select good build configurations. However, the model based on Bayesian optimization 
provides better prediction accuracy.
Further, \tool{} provides several insights that can be used by package managers as well as developers
to determine possible version conflicts as well as understand the sensitivity of the build outcome to
certain packages versions.

\section{Background}

The appeal of component software is clear; dividing software into logical components enables developers to separate concerns and reuse software more easily. Experts on one part of a system can independently develop a component package and expose an interface for other teams to rely on. Teams that rely on an external dependency can easily upgrade to newer versions with bug fixes and security updates, provided that the dependency's interface is still compatible.  Ensuring this compatibility, or at least reliably predicting it, is the challenge we aim to address in this paper.

\subsection{Versioning}

Package managers rely on version metadata to determine package compatibility. Ultimately, this metadata comes from humans (package developers and maintainers) with some knowledge of the package and its dependency relationships.
Developers can declare 
dependencies using a \emph{fixed} version (e.g. use only version $1.9.2$) or a 
\emph{version range} (e.g use a version which is at least as recent as $1.0.2$ or below version $1.9.2$ or between two $1.8.2-1.9.2$). 
While \emph{fixed} versions improve build determinism, they limits the
flexibility of a package. Supporting only a specific version can conflict with the versions required by other dependent packages, and this limits composability by making it very difficult to integrate packages into large projects.
The problem with flexible version ranges is that maintainers cannot test all possible configurations. Rather than exhaustively testing the versions in a range, developers more often simply assume that compatibility is consistent across all of them.

\subsubsection{\bf Balancing versions} Maintainers must strike a delicate balance. They need to keep their dependency specifications as flexible as possible to benefit from bug fixes and new features of dependent packages. However, they must also avoid including incompatible updates. At the same time, package developers must be careful not to release new versions with breaking changes.
Conventions such as semantic versioning, or \emph{semver}~\cite{preston2013semantic}, can help to 
identify new releases that either preserve or break
compatibility, but semantic versioning reliees heavily on developers to know the rules
of compatibility and versioning their package releases.
Such rules are complex and typically not fully understood~\cite{dietrich2016java} and, 
although developers are getting better with them, there is still a lot missing~\cite{decan2019package, dietrich2019dependency}. 
Tools that identify incompatible changes through static analysis are still incomplete~\cite{jezek2017api}, and
whereas testing~\cite{claessen2000quickcheck}
provides some coverage but cannot provide guarantees~\cite{hejderup2022can}. Because of this,
trying new dependency can still be a harrowing experience.
Even minor or patch version changes frequently break builds,
and finding a working configuration can involve manual
experimentation.

\subsubsection{\bf Version exploits}
Version flexibility has been exploited repeatedly. Most recently in the NPM ecosystem, a maintainer intentionally sabotaged a popular 
NPM package by removing it from NPM and deleting its repositories~\cite{npm_colors}.  After it was reinstated by the community and NPM,
the maintainer uploaded a version \emph{with a higher version number} introducing changes designed to break any software that used the package.
Dependent packages immediately started to fail, resulting in thousands of bug reports. Another exploit called {\it dependency confusion}~\cite{dependency_confusion} takes advantage of
version priority. A project with private dependencies 
meant to be fetched internally may also rely on a public
package repository for open source dependencies. Attackers can add an identically named, higher-versioned malicious package to an {\it external} repository and override the internal package, allowing them to run code on private systems.

\begin{figure}
     \centering
     \begin{subfigure}{0.45\textwidth}
         \includegraphics[width=\textwidth]{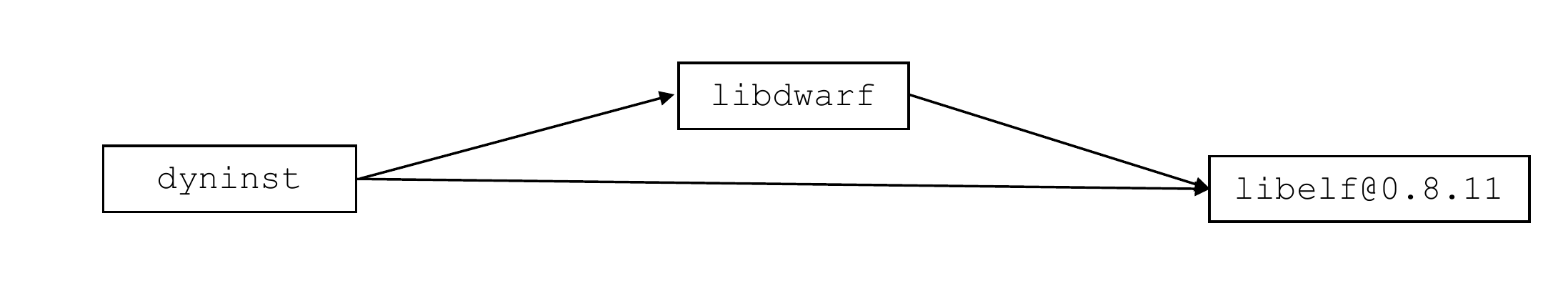}
         \caption{Abstract Spec: DAG representation of the spec:  ``\textit{dyninst \^libelf@0.8.11}'' . 
         The spec allows freedom to  the Spack concretizer to select any version for \textit{dyninst} and \textit{libdwarf} but constraints the \textit{libelf} package to use the fixed version \textit{0.8.11}}
         \label{fig:abstract_spec_dag}
     \end{subfigure}
     \begin{subfigure}{0.45\textwidth}
         \includegraphics[width=\textwidth]{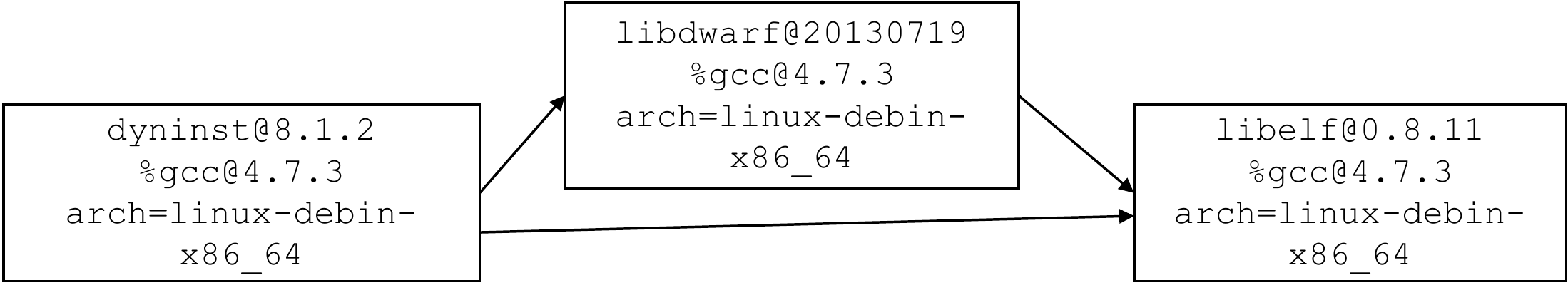}
         \caption{Spack provides as input to the concretizer the abstract spec depicted in Figure \ref{fig:abstract_spec_dag}. 
         The concretizer produces as output a concrete spec. Namely, in a concrete spec each node in the dag has all the information fully defined.}
         \label{fig:concrete_spec_dag}
     \end{subfigure}
     \caption{Abstract and concrete {\it spec} DAG representations. }
     \label{fig:spec_dag}
\end{figure}

\subsection{Spack}

In this paper we use the Spack~\cite{gamblin2015spack} package 
manager to explore the combinatorial build space of packages
and to understand what factors contribute to a successful build.
We chose Spack for several reasons. First, because 
it comes from the world of High Performance Computing~(HPC), Spack
is designed to support building packages from source.
Second, as HPC engineers frequently need to port packages to new 
systems, Spack is {\it very} flexible about versions, build 
parameters, and dependency configurations. It exposes
parameters to users as adjustable knobs and allows a single
package to be built in many different ways. This is ideal for 
experimentation.  Finally, like system-level Linux package 
managers, Spack is language-agnostic. Its focus is on HPC,
scientific computing, and
machine learning, so C, C++, Fortran, Python, and various
parallel programming models comprise most of the $>6,000$ packages
in its ecosystem. Scientific
software is notoriously difficult to build~\cite{gamblin2015spack,kumfert+:2002,hoste-easybuild-pyhpc-2012,dubois2003johnny}, and our aim is to simplify build configuration.

\begin{figure}
     \centering
     \includegraphics[width=\columnwidth]{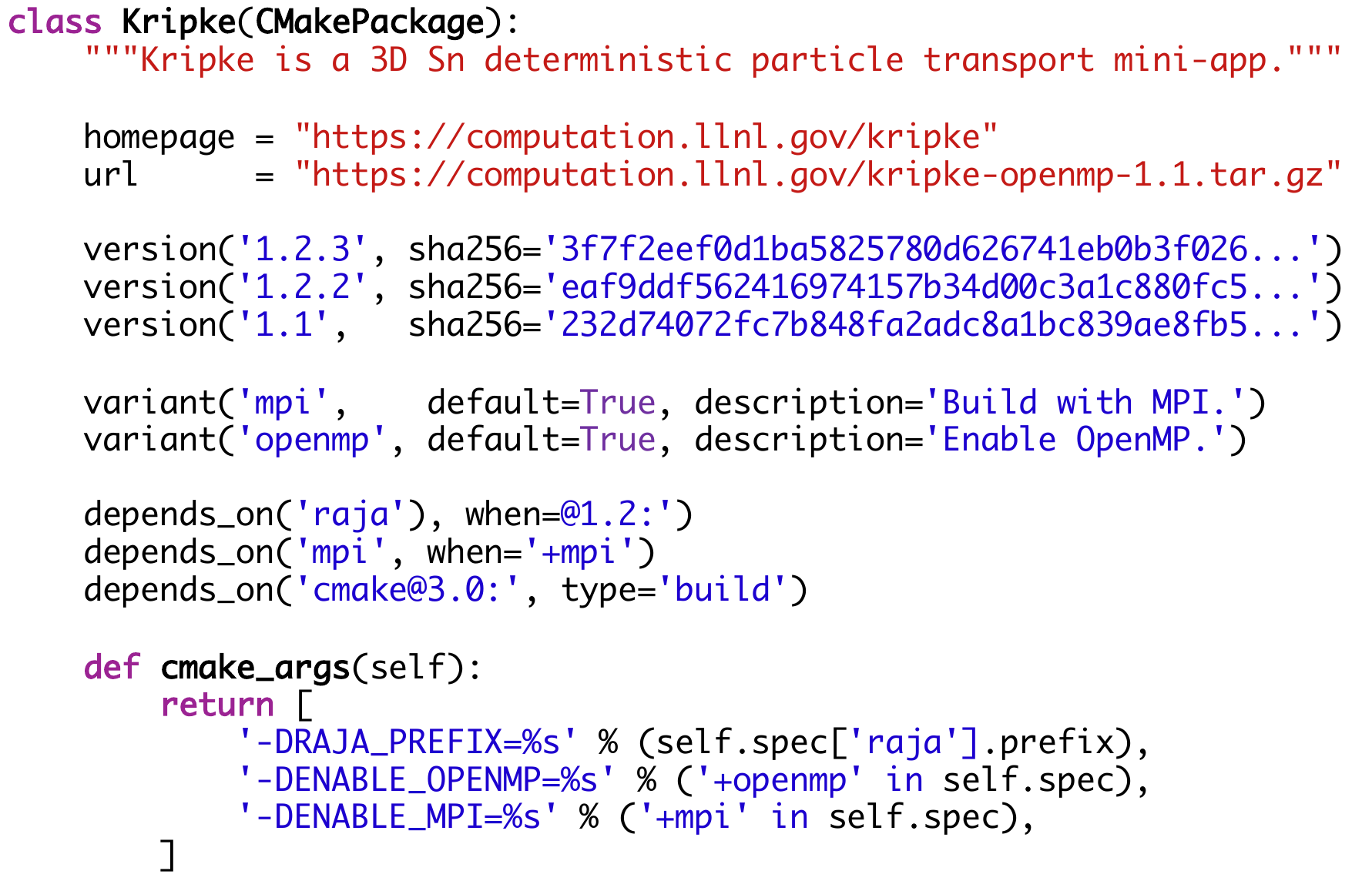}
     \caption{Example of a parameterized Spack package. }
     \label{fig:spack-package}
\end{figure}

\subsubsection{\textbf{Package DSL:}}
Spack packages are written in a domain-specific langage embedded in
Python, as shown in the example {\tt Kripke} package in 
Figure~\ref{fig:spack-package}. Each package is
a class containing {\it directives} at the class-level to constrain
the build space and {\it functions} that describe how to build. 
In the figure, {\tt version} directives describe available version
tarballs and their checksums, {\tt variant} directives expose
optional build parameters, and {\tt depends\_on} directives define
dependencies on other packages. Packages may require particular
versions of dependencies, e.g., {\tt kripke} requires {\tt cmake} at
version {\tt 3.0} or higher. Dependencies can also be conditional on
versions, variants, or other properties. Here, {\tt kripke} only
depends on {\tt raja} when it is at version {\tt 1.2} or higher, and
it only depends on {\tt mpi} when the {\tt mpi} variant is enabled.
The {\tt raja}, {\tt mpi}, and {\tt cmake} packages are described by
separate package templates like this one.

Directives are supplied by package maintainers to define the
combinatorial space of configuration that are {\it possible}. Spack 
selects a concrete configuration, or {\it spec}, from this space and
builds it using the recipe described in the package's functions. Here,
the package extends a base {\tt CMakePackage} class that has the bulk
of the standard recipe for handling the CMake build system, and
the {\tt cmake\_args} method tells the package how to translate
the {\tt spec} configuration to arguments for the {\tt cmake} command
that configures the build.

\subsubsection{\textbf{spec syntax:}}
The \emph{spec syntax} in Spack allows users to specify their own
constraints on top of those provided by maintainers in package
recipes. 
For example, often a user may only care to build a particular package with no additional constraints, or build
a package with a specific compiler. Table~\ref{tab:spack_syntax} presents some examples of the 
\verb|spack install| command, the spec syntax and a description. The spec syntax is fully recursive, in that any dependency can be constrained, just as the root node can. 

\begin{table}[h]
\caption{Examples of the \texttt{spack install} command using the \emph{spec} to constrain the combinatorial build space.}
\label{tab:spack_syntax}
\begin{tabular}{m{0.3\columnwidth}m{0.65\columnwidth}}
\toprule
\textbf{Examples} & \textbf{Meaning} \\
\midrule
\footnotesize \verb|spack install kripke| & Install {\tt kripke} but do not constrain the installation; Spack decides how to build. \\
\midrule
\footnotesize \verb|spack| \verb|install| \verb|kripke| \verb|%gcc@8.3.1| & Build {\tt kripke with {\tt gcc} 8.3.1} \\
\midrule
\footnotesize \verb|spack install| \verb|kripke@1.2.2~mpi| \verb|^raja@1.9| & Build {\tt kripke} version 1.2.2 with {\tt mpi} disabled and with {\tt raja} version 1.9. \\
\bottomrule
\end{tabular}
\end{table}

\subsubsection{\textbf{Concretization:}}
The specs in the table are {\it abstract} -- that is, they do not
specify {\it all} aspects of build configuration. Typically,
users care about a 
very small set of constraints, and rely on Spack to decide the rest
of the build configuration. In Spack, the process of selecting a consistent build configuration is called {\it concretization}.  In other package managers, the analogous process
is typically called {\it dependency resolution}. 
Figure~\ref{fig:abstract_spec_dag} presents an abstract spec DAG along with its corresponding {\it concrete} spec. The concrete graph is created by combining constraints from packages, command line, and user preferences and solving for what nodes and configuration options
should be present. Constraints not defined in the abstract spec represent degrees of freedom.

Spack's {\it concretizer} is a combinatorial logic solver implemented
using Answer Set Programming~(ASP)~\cite{gebser+:aicomm11}
It translates inputs from 
\begin{inparaenum}[i)]
\item the \emph{package-DSL},
\item the \emph{spec-syntax}, and
\item configuration file
\end{inparaenum} into a Prolog-like program that finds a
dependency graph satisfying all constraints from the
the \emph{package DSL} and the \emph{spec syntax} on the command line.
The concretizer eases the task of exploration because it ensures
that only valid configurations are resolved.
However, there can exist multiple
configurations which satisfy those constraints. Typically,
users must explore this space manually by specifying constraints,
but our goal in this work is to automatically find good configurations {\it within} the set of valid configurations.

\subsection{\textbf{Vision for the Future}}

In an ideal software ecosystem the package authors and maintainers would be able to define dependencies without any specific constraints and fully embrace flexible/range versioning schemes.
The package managers of the future should be able to automatically select compatible configurations that successfully build. 
Spack provides rich abstractions to maintainers and users to fully 
embrace such a software ecosystem. 
In essence, Spack separates concerns. It allows packages to be extremely flexible while
also providing syntax to the user to fix constraints. However, currently this design choice burdens the user. Since they need 
to identify configurations that will build from a potentially sparse space of options.
Our approach tries to bridge this gap.
\section{High-Fidelity Build Configuration Selection}
Identifying compatible versions of a package's dependencies is a time consuming and error-prone endeavour. 
Currently, the process of identifying a package configuration that successfully builds
relies on  individuals with in-depth knowledge of various 
packages and their interactions with their dependencies. In the absence
of such knowledge users have to resort to exploring the build configuration space by
trying out various combinations of packages and their dependencies. The space
formed by all possible options for a package and dependencies is immense, and as a result 
an exhaustive search of that space is impractical. One can resort to random sampling,
but that is still a tedious process, and if system parameters change it would
have to be redone all over again. Moreover, just building the package itself takes a significant amount of time.
Ideally, we want an automated process that can select select configurations that are highly likely to build based on a smaller set
of samples.

The trick to learning quickly is to pick samples carefully, and we develop an active-learning-based approach to identifying high-fidelity package build configurations using only a limited set of samples. Active learning
%
%
algorithms select data from which to learn in order to achieve
a specific objective. This is especially suitable when the true objective function evaluations 
are expensive, as is in the case of building packages. Below, we describe the problem formally
and provide details about the iterative sampling, design of the model, and describe the 
selection algorithm in its entirety. 

Let the root package we are building require $n$ dependencies, each of which
is represented by $X_i, i \in{1,\dots,n}$. A package $X_i$ can take on values $x_i \in {v_{i,1}, \dots, v_{i,m_i}}$.
We use $ \bm{x} \equiv [x_1,\dots,x_n] \in \mathcal X$ to represent a
configuration of the $n$ dependent packages as a vector. Let $f:\mathcal X \rightarrow \{0, 1\}$ be a function representing the outcome of building a configuration.
If a configuration $\xx$  builds successfully, then $f(\xx)=1$ and if it fails to build, then $f(\bm{x})=0$.
Our goal is to find \emph{some} configuration $\xx^* \in \mathcal{X}$, from the space of all possible configurations
$\mathcal{X}$, that successfully builds. This can be represented by the following objective function.
\begin{equation*}
    \xx^* = \argmax_{\xx} f(\xx)
\end{equation*}
Note that, while there may be many configurations that successfully build, we are interested in finding any one of those.





\subsection{Iterative Configuration Selection Algorithm}\label{sec:sampling}
We use an iterative process for sampling configurations that are likely to build.
A probabilistic surrogate model is constructed that predicts the probability that
a configuration will successfully compile. A
set of highly likely candidates is sampled and built. The outcome is then used to update the model and the process is repeated. Thus, our method follows an adaptive iterative scheme, where it alternates between updating the model and using it to make choices 
about which configurations to investigate next. Our method is sketched in Algorithm~\ref{alg:sampling}.

\begin{algorithm}[H]

\caption{Pseudocode for Adaptive Sampling}
\label{alg:sampling}
\begin{algorithmic}[1]
\Procedure{Sample} {$f$,$\MM_0$,$\mathcal{S}$,$T$}

\State $\HH_0 \leftarrow \emptyset$
\For{$t=1$ to $T$}
\State $\xx_t^* \leftarrow  \argmax_{\xx} \MM_{t-1}(\xx)$
\State $y^*_t \leftarrow f(\xx_t^*)$ \Comment{Expensive process}
\State $\HH_t \leftarrow \HH_{t-1} \cup \{(\xx^*_t,y^*_t)\}$
\State $\MM_t \leftarrow \text{FITMODEL}(\MM_{t-1}, \HH_t)$
\EndFor
\State \textbf{return} $\HH_T$
\EndProcedure
\end{algorithmic}
\end{algorithm}

In each iteration $t$, a single configuration $\xx^*_t$ is selected, from a list of candidate configurations.
As evaluating the true objective function for all the candidate configurations is expensive, we use a surrogate model $\MM_{t-1}$ to score the candidates.
The surrogate model is constructed from the observation history, $\HH_{t-1}$, and is cheap to evaluate. The candidate  $\xx^*_t$ with the maximum score value is selected at each iteration and is use to evaluate the true objective
function, $y^*_t = f(\xx^*_t)$, by running the build process for the configuration $\xx^*_t$. The 
result of the evaluation is then added to the observation history $\HH_{t-1}$ to yield the updated history $\HH_t$. Finally, $\HH_t$ is used to update
the model, yielding $\MM_{t}$, and choose the next configuration $\xx^*_{t+1}$ at iteration $t+1$. This process continuous iteratively until iteration $t=T$. 
The algorithm starts off with a small set of $20$ initial samples drawn
uniformly at random. All of them are evaluated to obtain initial observation history $H_0$ and then create the initial surrogate model $\MM_{0}$.
This method is a form of Sequential Model Based Global-Optimization methods~(SMBO), and has been 
used for black box optimization of functions that are expensive to evaluate~\cite{hutter2011sequential,jones2001taxonomy,villemonteix2009informational}.
\subsection{Probabilistic Surrogate Model}
A surrogate model is a model that approximates the true objective function, $y=f(\bm{x})$, and provides
a significantly cheaper method for computing the approximate objective for all samples. 
Given a space $\mathcal{X}$ of all possible build configurations, our goal is to design a surrogate model that provides a score indicating whether a configuration $\bm{x} \in \mathcal{X}$ is likely to build
A good surrogate model will assign
a high score to each good configuration. As the surrogate is significantly cheaper
to evaluate than the true objective function, the iterative
sampling algorithm (see section~\ref{sec:sampling}) can use it to rank all the configurations
and select the most promising configuration whose true objective function can then be evaluated
by running the build process. 
We describe two surrogate models that are incorporated in \tool{}. 

\subsubsection{Bayesian Model}
To construct this model, we use the algorithm used in Bayesian optimization, which has been used for
tuning the hyperparameters of deep neural networks~\cite{bergstra2011algorithms}, where training is a very expensive
process. 
The surrogate model $\MM(x)$ computes Expected Improvement~($\EI$)~\cite{jones2001taxonomy} using a probabilistic model $p_{y\mid \xx}(y\mid \xx)$ for some configuration $\xx$.
The Expected Improvement $\EI$~(\cref{eq:eipx}) is the expected margin by which the true objective $f(\xx)$ will be $1$ (successfully builds).

In the Bayesian optimization method, Bayes rule is used
to define $p_{y\mid \xx}(y\mid \xx)$ in terms of $p_{\xx \mid y}(\xx \mid y)$, yielding: 
\begin{equation}
p_{y\mid \xx}(y\mid \xx) = \frac{p_{\xx \mid y}(\xx \mid y) p_y(y)}{p_{\xx}(\xx)}
\label{eq:bayes}
\end{equation}
Here, $p_y(y)$ is the prior distribution for $y$ and
$p_{\xx}(\xx) = \int p_{\xx \mid y}(\xx \mid y) p_y(y) dy$ is the marginal distribution of $\xx$.

To model $p_{\xx \mid y}(\xx \mid y)$ we split $y$
into two possibilities: 1) successfully builds i.e. $y = 1$, and 2) 
fails to build i.e. $y=0$.
This enables us to define the probability distribution $p_{\xx \mid y}(\xx \mid y)$ in
terms of two probability density functions; $\PG(\xx)$ for good configurations, and $\PB(\xx)$ for bad configurations:
\begin{equation} 
p_{\xx \mid y}(\xx \mid y) = \begin{cases}
  \PG(\xx) \quad \text{if $y =1$}\\
  \PB(\xx) \quad \text{if $y =0$},
  \end{cases}
\label{eq:pxy}
\end{equation} 

Finally, the Expected Improvement $\EI(\xx)$ can now be written as:
\begin{align}
\EI(\xx) & = \frac{1}{\alpha + \frac{\PB(\xx)}{\PG(\xx)} (1-\alpha)}
\label{eq:eipx}
\end{align}
We can compute $\EI(\xx)$ for each candidate configuration and choose the one with the
highest value as the candidate $\xx_t^*$ that is most likely to build. $\xx_t^*$ can
then be used for performing the full evaluation $y_t^* = f(\xx^*_t)$.

To construct $\PG(\xx)$ and $\PB(\xx)$, we leverage the underlying dependency structure of the package.
We represent a package and its dependencies using an undirected graph, and consider a joint distribution 
over the dependencies to calculate the score. Let $G=(X,E)$ denote the undirected graph corresponding
to the root package of interest, where $X$ is the set containing the package and the dependencies and $E$ is
the set of edges as given by the dependency structure of the package. 

Our surrogate model uses a probability density defined over the prediction values $y$ given
$\bm{x}$. Estimating the full joint distribution over the parameter space is not feasible. 
However, we can leverage the inherent relationship between the packages provided by the package dependency graph
to obtain a factorized distribution. This allows us to represent the joint probability distributions as a
product of factors with smaller number of variables. Let the factors be $\{F_j(V_j)\}_m$, where
$V_j \subseteq \{X_1, \dots, X_n\}$ is a subset of the variables. Then we can write the joint distribution $\PG(\xx)$ as
\begin{align}
    \PG(\xx) &= \prod_{j=1}^m F^{(g)}_j(V_j)
    \label{eq:p_fact}
\end{align}
We use the $(g)$ superscript to denote the factors corresponding to distribution for the good configurations $\PG$. We consider factors of at most size two, that are defined for every node and edge of the dependency graph. For example, a factor $F^{(g)}(\{X_j,X_k\})$ corresponding to an edge between the node $j$ and the node $k$ captures the likelihood of that pair compiling. 

The underlying assumption behind this factorization is that the probability of a package building
will depend on whether that package and its dependencies can be built successfully. In the event where a package and
its immediate dependency fails to build, the package fails to build because its requirements cannot be satisfied. Hence, instead of
estimating the full joint distribution, we can reasonably approximate
the probability of a package building as a factorization over the probability distribution of the pairwise distributions between each parent and child (given by
the dependency graph).
$\PB(\xx)$ is constructed similarly to Eq~\ref{eq:p_fact} using the same factorization $\{V_j\}_m$ but a different set of factor functions $F^{(b)}_j$.

\subsubsection{Wisdom of the Crowd}
This approach uses the majority opinion to select versions for 
packages~\cite{mileva2009mining}. The probability of selecting a version for a package is directly
proportional to the number of times it occurred in successfully built configurations in the observed data.
This can be represented as a probability distribution as shown below.
\begin{align}
    p(\bm{x}) &= \prod_{i=1}^n \PG(v_i)
    \label{eq:p_wisc}
\end{align}
where $\PG(v_i)$ is the probability of finding package $v_i$ of $X_i$
in the observed set of good configurations.



\subsection{Pairwise Importance Analysis}
A particular choice of version for a package or it's dependencies can have a significant impact
on whether that package builds. However, not all parent and child pairs in the package dependency tree will have 
similar influence on the final outcome.  Some package pairs might be compatible across all possible 
versions and some might have very strict compatibility rules. Identifying the pairs that are crucial for
the success or the failure of the build process enables users to better understand the most likely sources of failure. 
We use the two probability densities from the surrogate model to derive a metric
that tells us the relative significance of each package in
build outcomes. Recall that the surrogate model maintains two distributions: a distribution
of package versions based on high-fidelity configurations $\PG$, and a distribution of package
versions based on bad configurations $\PB$. The degree to which these distributions
differ from one another is an indication of how sensitive the build outcome is
to the version of this package. We use Jensen-Shannon~(JS) divergence to 
compute the difference between the two distributions. 

For two probability distributions $P$ and $Q$ defined in the same probability space
$X$, the JS divergence is defined using $M = (P+Q)/2$ as
\begin{align}
D_{JS}(P,Q) &= \frac{1}{2}D_{KL}(P, M) + \frac{1}{2}D_{KL}(Q, M) \\
D_{KL}(P,M) &= \sum_{x\in X} P(x) \log \frac{P(x)}{M(x)} 
\label{eq:d_js}
\end{align}
Here $D_{KL}(P,M)$ is the Kullback--Leibler (KL) divergence from $M$ to $P$. $D_{KL}(Q,M)$ is defined similarly. Note that $D_{JS}(P,Q) \ge 0$, with equality for identical distributions.

\section{Experimental Setup}
In this section we provide details about the dataset used, data collection mechanism, and the
metrics used for our evaluation.

\subsection{Evaluation Dataset}
\begin{table*}[ht]
\caption{The number of tested configuration for various packages within the E4S software ecosystem. In total we
explore $31,186$ unique root package builds resulting in examining $83,796,782$ root packages and dependencies.}
\label{tab:package_details}
\footnotesize
\begin{tabular}{lccc lccc lccc}
\toprule
 Package & Configs(\#) & Good (\#) & Deps  (\#) &  Package & Configs(\#) & Good (\#) & Deps  (\#) &  Package & Configs(\#) & Good (\#) & Deps  (\#) \\
\cmidrule(lr){1-4} \cmidrule(lr){5-8} \cmidrule(lr){9-12} 
abyss	&	892	&	133	&	36	&adios	&	58	&	11	&	48	&amrex	&	616	&	543	&	40	 \\
ascent	&	320	&	14	&	32	&axom	&	166	&	5	&	40	&bolt	&	996	&	53	&	22	 \\
cabana	&	428	&	130	&	42	&caliper	&	289	&	221	&	41	&chai	&	298	&	19	&	16	 \\
conduit	&	599	&	241	&	29	&darshan-runtime	&	190	&	163	&	39	&hypre	&	322	&	264	&	40	 \\
hpx	&	799	&	384	&	45	&heffte	&	292	&	163	&	33	&hdf5	&	691	&	641	&	40	 \\
gmp	&	636	&	18	&	19	&globalarrays	&	323	&	262	&	40	&fortrilinos	&	146	&	8	&	30	 \\
faodel	&	361	&	269	&	29	&kokkos	&	625	&	294	&	13	&kokkos-kernels	&	567	&	115	&	14	 \\
libnrm	&	923	&	48	&	40	&mercury	&	990	&	539	&	49	&metall	&	275	&	261	&	14	 \\
mfem	&	441	&	348	&	119	&mpark-variant	&	198	&	126	&	13	&ninja	&	621	&	503	&	23	 \\
omega-h	&	999	&	213	&	42	&openmpi	&	333	&	12	&	114	&openpmd-api	&	90	&	28	&	41	 \\
papyrus	&	977	&	672	&	40	&parallel-netcdf	&	345	&	25	&	39	&petsc	&	403	&	26	&	141	 \\
phist	&	852	&	434	&	183	&plasma	&	993	&	378	&	14	&pumi	&	990	&	914	&	40	 \\
py-libensemble	&	381	&	18	&	56	&py-petsc4py	&	119	&	7	&	50	&qt	&	670	&	6	&	177	 \\
qthreads	&	94	&	94	&	22	&qwt	&	87	&	21	&	102	&raja	&	82	&	46	&	15	 \\
rempi	&	848	&	135	&	39	&scr	&	30	&	20	&	65	&slate	&	169	&	97	&	44	 \\
slepc	&	126	&	33	&	45	&stc	&	692	&	3	&	41	&sundials	&	708	&	636	&	40	 \\
superlu-dist	&	624	&	58	&	43	&swig	&	215	&	183	&	18	&sz	&	652	&	497	&	14	 \\
tasmanian	&	860	&	709	&	40	&trilinos	&	1000	&	168	&	65	&turbine	&	565	&	369	&	34	 \\
umap	&	702	&	610	&	13	&umpire	&	82	&	52	&	15	&unifyfs	&	246	&	13	&	41	 \\
upcxx	&	994	&	539	&	34	&variorum	&	989	&	288	&	24	&veloc	&	667	&	51	&	37	 \\
zfp	&	540	&	336	&	13	 \\

\bottomrule
\end{tabular}
\end{table*}

We evaluated our model on several packages from the Extreme-scale Scientific Software Stack (E4S).
E4S provides open source software packages for developing, deploying and running scientific 
applications on high-performance computing (HPC) platforms. These software packages are implemented in
different programming languages such as C/C++, FORTRAN,  Python, Lua, and others.
E4S uses Spack for managing
software packages. Table~\ref{tab:package_details} shows the list of packages from E4S that were used for
our evaluation.

\subsection{Data Collection}

We explored the build space for several Spack packages from the E4S software stack. The combinatorial package space consists of millions of configurations. So, to create the dataset we randomly select a set of concrete specs from that space. We refer to these concrete specs as configurations. The selection samples only different versions and does not sample variants or compiler options. 

We evaluate whether each configuration successfully builds or not. Initially, we create a DAG in which the nodes are the unique packages of all the sampled configurations. Common dependencies among configurations appear only once in the DAG and, thus, we built those only once. We implement a parallel build process of the entire DAG in a distributed system. The process follows a \emph{farmer-worker} parallel paradigm. A single node operates as the farmer. The farmer traverses the DAG and selects the nodes ready for installation. A node is ready when all the dependencies of the node have been successfully built. Once the farmer gathers these nodes, it assigns their installation to workers. The workers operate in parallel within the distributed cluster and issue the \verb|spack install "specification"| command to build the individual nodes. Once all workers finish their installations, the farmer identifies which installations failed or succeeded and updates the DAG. When a dependency fails, all dependent nodes are marked recursively as \texttt{failed}. The process continues until all nodes in the DAG are marked as failed or successful. In the end, the farmer exports the DAG information and the node installation status to a pandas~\cite{pandas} \textit{DataFrame} and stores it into permanent storage. 

The farmer-worker installation protocol relies on Spack being parallelism aware and allowing multiple Spack instances to execute concurrently without corrupting internal Spack data structures. Spack implements a locking mechanism through the network filesystem protocols and uses the lock when updating internal data structures (such as the user spack database). During our evaluation, we repeatedly pushed the locking mechanism to its limits. The process exposed concurrency bugs in the mutual exclusion
algorithm implemented in Spack. We communicated and fixed those bugs resulting in a significantly improved locking mechanism.  In the end, the farmer-worker protocol coupled with the improved locking mechanism resulted in executing up to $432$ concurrent spack instances with each instance using 9 CPU cores.

In Table~\ref{tab:package_details} we show the total number of configurations for each package, the number of successful builds, and the number of dependency packages. Naively, building this data set would require $83,796,782$ package installations, but Spack identifies unique builds with a Merkle hash of their configuration metadata. That is, the uniqueness of a {\it configuration} depends on the root's metadata {\it and} the metadata of all of its dependencies. By hash, there were $56,645$
unique package configurations (including root packages and dependencies). We skipped the installation of $19,662$ nodes due to a failed
%
%
dependency build, and we performed $31,186$ total package installations out of which $28,157$ were successful. 
This dataset was collected on a $3,018$-node Intel Xeon cluster with $36$ cores per node.

\begin{figure*}[ht]
\centering
\includegraphics[width=0.9\textwidth]{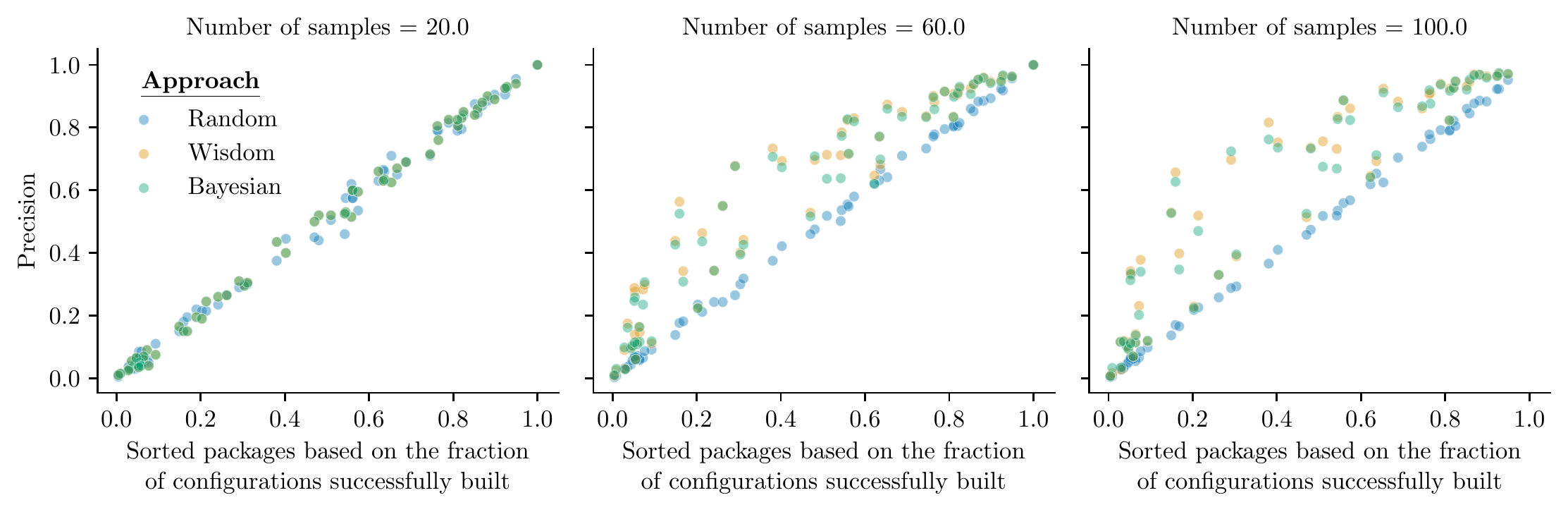}
\caption{Precision for all the packages. Each point on these scatter plots corresponds to a package. 
Both the models of \tool{} (\emph{Bayesian} and \emph{Wisdom of the crowd})
have a significantly higher precision than \emph{Random Selection}. \tool{} is
bootstrapped with $20$ samples drawn uniformly at random, so performance of
all the approaches is similar for sample size $20$.}
\label{fig:precision_all}
\end{figure*}

\begin{figure*}[ht]
\centering
\includegraphics[width=0.9\textwidth]{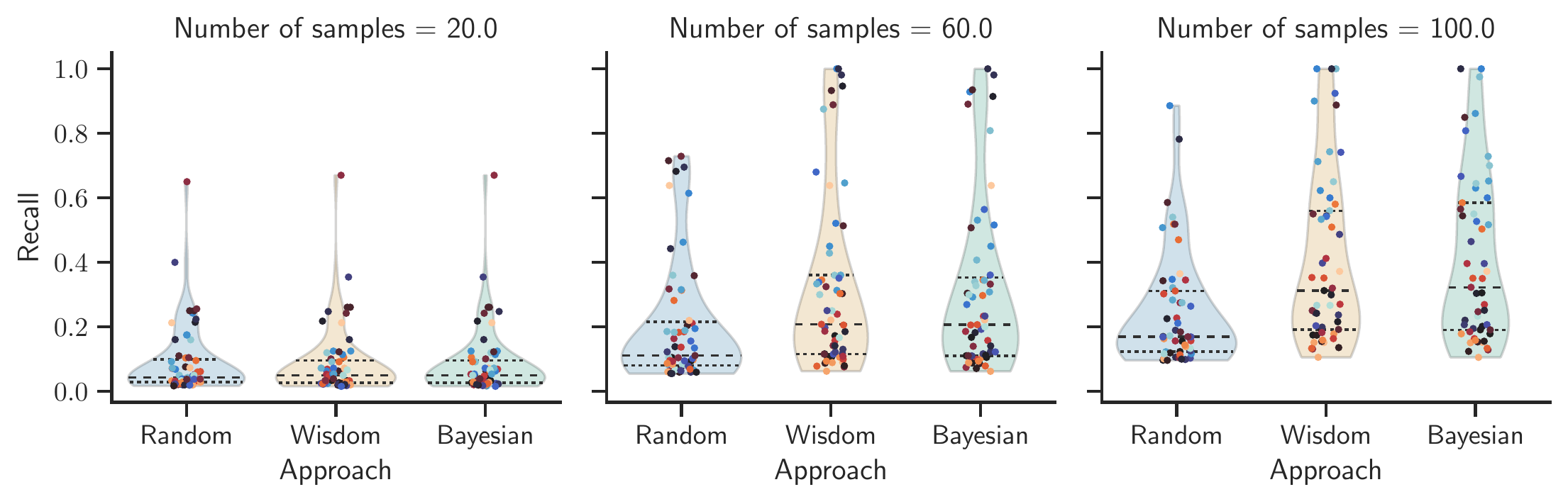}
\caption{Recall for all the packages. The violin plots show the probability density of recall values along with
the interquartiles means for different sample sizes. Each point shown here corresponds to a package color-coded
based on their overall build success rates (light blue to dark red corresponds to build success rates ranging from $0$ to $1$). 
$\mathcal{R}$ value of $1$ indicates that all the good configurations are included
in the selected samples. Both the models of \tool{} (\emph{Bayesian} and \emph{Wisdom of the crowd})
have a higher recall score than \emph{Random}, and for some packages attains $\mathcal{R}$ of $1$. }
\label{fig:recall_all}
\end{figure*}

\subsection{Metrics for Evaluation}
A good sampling selection algorithm is expected to identify high fidelity configurations while 
observing fewer samples. As a result we use the following metrics to evaluate our approach.

\noindent \textbf{Precision ($\mathcal{P}$)} tells us what fraction of the samples contain configurations that
built successfully.
    \begin{align}
    \mathcal{P(H)}     = \frac{|\{x|x\in \mathcal{H}, f(x) = 1\}|}{|\mathcal{H}|}
\end{align}

\noindent \textbf{Recall ($\mathcal{R}$)} gives the ratio of the configurations that successfully built and included
in the sampled set to the actual number of configurations that successfully built in the entire sample
space. 
\begin{align}
    \mathcal{R(H)} = \frac{|\{x|x\in \mathcal{H}, f(x) = 1\}|}{|\{x|\forall x, f(x) = 1\}|}
\end{align}

\noindent\textbf{Area Under the Precision-Recall Curve (AUPRC):} 
A PR curve shows the trade-off 
between precision ($\mathcal{P}$) and recall ($\mathcal{R}$) across different sample sizes. 
AUPRC is obtained by calculating the area under the PR curve, and is calculated as shown below.
\begin{align}
    AUPRC = \sum\limits_{k=1}\limits^N \mathcal{P}(\mathcal{H}_k) \Delta\mathcal{R}(\mathcal{H}_k)   \label{eq:auprc}
\end{align}
where $N$ is the total number of configurations in the collection, $\mathcal{P}(\mathcal{H}_k)$ is the precision 
for $k$ samples, and $\Delta \mathcal{R}(\mathcal{H}_k)$ is the change in recall that happened between $k-1$ and
$k$ samples.
The AUPRC is a useful metric in problem settings like this where we care about finding positive samples.
In our case, AUPRC will provide a measure of how well the surrogate model is able to select
high-fidelity build configurations. AUPRC value of $1$ indicates a model that is able to
select high-fidelity build configurations perfectly.

\section{Evaluation}
In this section we evaluate our approach on the datasets listed in Table~\ref{tab:package_details}. For each package,
we compare the performance of all the methods for a range of samples by running the 
algorithm $10$ times and reporting the mean for each evaluation metric. 
We evaluate \tool{} by comparing the Bayesian model and the Wisdom model against Random Selection.
In \emph{Random Selection}, the configurations are selected uniformly
at random from the database.



\subsection{Evaluation of Build Configuration Selection}
A package dataset consists
of several build configurations, of which only a subset successfully built. The goal of configuration selection
is to identify high-fidelity builds using fewer samples.
Figure~\ref{fig:precision_all} shows the $\mathcal{P}$ for different packages at different
sample sizes. The x-axis lists the packages sorted based on their build success rate (number of
good configurations divided by the total number of configurations).
The figure shows that both \tool{} models (\emph{Bayesian} and \emph{Wisdom of the crowd})
have a significantly higher fraction of high-fidelity build configurations
compared to \emph{Random Selection}.
As mentioned in section~\ref{sec:sampling}, we bootstrap our
search with an initial set of $20$ samples chosen uniformly at random, which is why 
all the algorithms have a similar performance for the sample size of $20$.
Figure~\ref{fig:recall_all} shows the $\mathcal{R}$ for different packages at different
sample sizes. The $\mathcal{R}$ metric for \emph{Bayesian} and \emph{Wisdom of the crowd} is
much higher than \emph{Random Selection} indicating that these approaches identify several, if not all, high-fidelity build configurations with far
fewer samples. Note that for packages with a higher build success rate, $\mathcal{P}$ would be close
to $1$, however $\mathcal{R}$ of $1$ can only be attained when at least that many
samples are selected. For some packages, our approach achieves an $\mathcal{R}$ of $1$ with just
$60$ or $100$ samples, indicating that \tool{} selected all the high-fidelity
builds. 




\subsection{Evaluation of the Models}
\begin{figure}
    \centering
    \includegraphics[width=0.8\columnwidth]{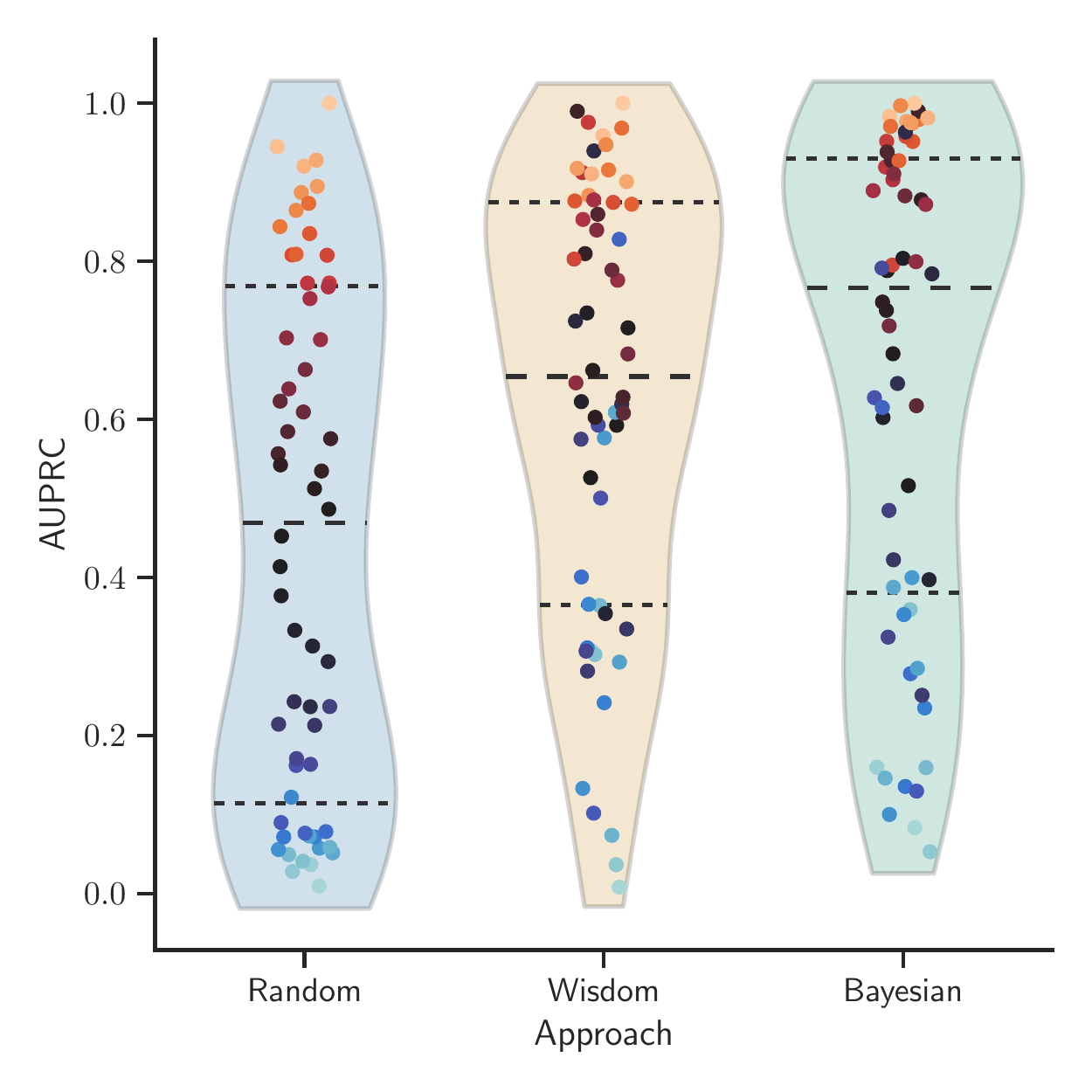}
    \caption{AUPRC metric for all packages. Each point represents a package and the
    color is assigned based on their build success rate (light blue to dark red corresponds 
    to success rates ranging from $0$ to $1$). This violin plot shows the probability density of the
    data at different AUPRC values. The \emph{Bayesian} model has a higher AUPRC
    value than \emph{Wisdom of the crowd} model indicating that \emph{Bayesian}
    is a better model for identifying all the high-fidelity build configurations.}
    \label{fig:auprc_all}
\end{figure}

\begin{figure}
\centering
\includegraphics[width=\columnwidth]{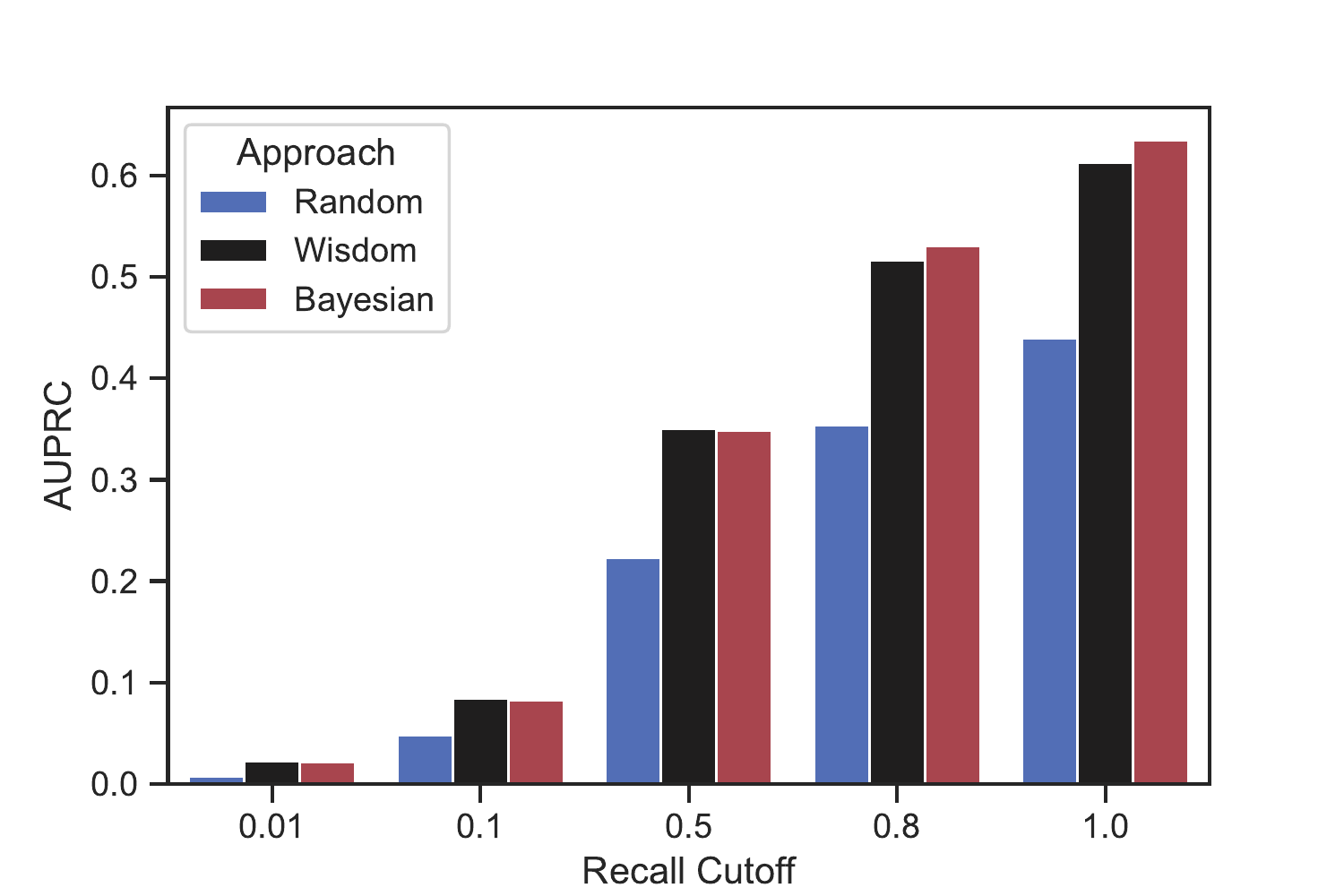}
\caption{Mean AUPRC metric for all packages at different recall cutoffs. While both
the models from \tool{} performs similar for small cutoffs, \emph{Bayesian} performs
better for larger cutoffs indicating that it better at selecting all the high-fidelity builds.}
\label{fig:auprc_cutoff}
\end{figure}

While the metrics $\mathcal{P}$ and $\mathcal{R}$ show us the ability
of the models to select a sample that successfully builds, they don't tell us how
well the model will perform if it has to identify all the good configurations. We use the AUPRC metric from
Eq~\ref{eq:auprc}, which calculates the area under the Precision Recall curve,
to compare the different algorithms. The ideal value for AUPRC is $1$ and the higher the value the better the
algorithm is at selecting good configurations. We divide the data into test and training
sets, train the models on the training set and use it to evaluate the test set. Half of the
dataset was used for training and it was tested on the other half. The model used the training set
and iteratively selected samples and updated the model. $100$ samples were selected.
We used this model to evaluate the test set and calculated the AUPRC metric.
Figure~\ref{fig:auprc_all} is a violin plot showing the AUPRC metric
for all packages in our dataset. It also shows the probability density of the data at different AUPRC values
as well as the interquartile range.
It can be seen that the \emph{Bayesian} model has a higher AUPRC
value than the \emph{Wisdom of the crowd} model, which indicates that \emph{Bayesian} is better at
selecting high fidelity build configurations. The figure also shows the distribution of AUPRC
values for different packages, where each point represents a package
color-coded based on their build success rates (light blue to dark red corresponds to build success
rates ranging from $0$ to $1$). The packages with higher build success rates tend to have
higher AUPRC values because they are relatively easier to identify and even \emph{Random Sampling}
can perform fairly well. For packages with lower build success rates (indicating fewer high-fidelity
builds in the build configuration samples), \emph{Bayesian} and \emph{Wisdoom of the crowd} is able
to achieve a higher AUPRC than \emph{Random Sampling} indicating that they are able to
select high-fidelity builds in tough cases.
We also show the mean AUPRC for different
recall cutoffs in figure~\ref{fig:auprc_cutoff}. While both the models from \tool{}
have similar performance for smaller recall cutoff, \emph{Bayesian} performs better with a larger recall cutoff, 
indicating that it is able to  more easily identify all the high-fidelity builds.

\subsection{Package Importance Analysis}
\begin{table*}[ht]
\caption{Relative ranking of dependencies for different root packages based on their importance. A parent child
pair is represented by \emph{parent}+\emph{child} as in the case of autoconf+m4.}
\label{tab:pkg_imp}
\footnotesize
\begin{tabular}{l|lllll}
\toprule
 Root package & Dependency ranking \\
 \hline
abyss & autoconf: 0.37& autoconf+m4: 0.37& autoconf+perl: 0.37& libtool+autoconf: 0.29& abyss+autoconf: 0.27\\
adios & autoconf+perl: 0.27& autoconf+m4: 0.27& autoconf: 0.27& libtool: 0.22& libtool+m4: 0.22\\
ascent & vtk-h+openmpi: 0.14& vtk-h: 0.14& vtk-h+vtk-m: 0.14& conduit+zlib: 0.12& conduit+hdf5: 0.12\\
axom & lua: 0.08& lua+ncurses: 0.08& lua+readline: 0.08& lua+unzip: 0.08& axom+openmpi: 0.07\\
bolt & autoconf+perl: 0.37& autoconf+m4: 0.37& autoconf: 0.37& automake+autoconf: 0.32& automake+perl: 0.30\\
hypre & openblas+perl: 0.07& openblas: 0.07& hypre+openblas: 0.03& hypre+mpich: 0.02& mpich+findutils: 0.01\\
hpx & hpx+boost: 0.24& hpx+hwloc: 0.24& hpx+pkgconf: 0.24& hpx+python: 0.24& hpx: 0.24\\
heffte & heffte: 0.35& heffte+openmpi: 0.30& heffte+fftw: 0.24& cuda+libxml2: 0.19& mpich+findutils: 0.19\\
hdf5 & mpich+findutils: 0.03& mpich+pkgconf: 0.03& mpich+libxml2: 0.03& mpich: 0.03& mpich+libpciaccess: 0.03\\
ninja & ninja+python: 0.03& python+ncurses: 0.01& python+readline: 0.01& python+pkgconf: 0.01& python+libffi: 0.01\\
omega-h & omega-h+zlib: 0.24& trilinos: 0.24& trilinos+openblas: 0.24& omega-h: 0.24& omega-h+trilinos: 0.18\\
openmpi & json-c: 0.30& mariadb+lz4: 0.30& meson: 0.30& gmp: 0.30& python+libffi: 0.30\\
openpmd-api & hdf5: 0.19& hdf5+zlib: 0.19& hdf5+openmpi: 0.19& hdf5+pkgconf: 0.19& hdf5+cmake: 0.19\\
papyrus & papyrus+mpich: 0.11& cmake+ncurses: 0.08& cmake: 0.08& papyrus+cmake: 0.08& mpich+findutils: 0.04\\
plasma & plasma: 0.52& plasma+openblas: 0.26& openblas+perl: 0.13& openblas: 0.13& plasma+cmake: 0.12\\
pumi & pumi+mpich: 0.02& mpich+findutils: 0.02& mpich+libxml2: 0.02& mpich: 0.02& mpich+libpciaccess: 0.02\\
py-petsc4py & py-mpi4py+python: 0.11& hypre: 0.11& hypre+openmpi: 0.11& py-mpi4py+py-setuptools: 0.11& hypre+openblas: 0.11\\
qthreads & numactl+libtool: 0.00& numactl+autoconf: 0.00& bzip2+diffutils: 0.00& util-macros: 0.00& hwloc: 0.00\\
raja & raja: 0.17& raja+cmake: 0.17& blt: 0.08& blt+cmake: 0.08& raja+blt: 0.07\\
rempi & autoconf+perl: 0.52& autoconf+m4: 0.52& autoconf: 0.52& rempi+autoconf: 0.50& automake+autoconf: 0.35\\
scr & libyogrt: 0.08& libyogrt+slurm: 0.08& scr+libyogrt: 0.07& scr+cmake: 0.05& scr+dtcmp: 0.05\\
slepc & hypre: 0.42& hypre+openmpi: 0.42& hypre+openblas: 0.42& superlu-dist+cmake: 0.39& superlu-dist: 0.39\\
superlu-dist & openblas+perl: 0.19& openblas: 0.19& cmake+ncurses: 0.11& cmake: 0.11& metis+cmake: 0.11\\
sz & cmake: 0.12& cmake+openssl: 0.12& cmake+ncurses: 0.12& sz+cmake: 0.04& sz+zstd: 0.03\\
trilinos & trilinos: 0.19& py-setuptools+python: 0.09& py-setuptools: 0.09& trilinos+mpich: 0.08& cmake+ncurses: 0.08\\
upcxx & upcxx: 0.24& upcxx+mpich: 0.22& upcxx+python: 0.11& mpich+findutils: 0.08& mpich+libxml2: 0.08\\
variorum & hwloc: 0.38& hwloc+ncurses: 0.38& hwloc+libpciaccess: 0.38& hwloc+pkgconf: 0.38& hwloc+libxml2: 0.38\\
veloc & veloc: 0.58& veloc+openmpi: 0.58& veloc+cmake: 0.58& veloc+openssl: 0.58& veloc+libpthread-stubs: 0.47\\
zfp & zfp+cmake: 0.06& cmake: 0.04& cmake+openssl: 0.04& cmake+ncurses: 0.04& bzip2+diffutils: 0.00\\
\bottomrule
\end{tabular}
\end{table*}

A particular choice of version for packages can significantly affect
the build outcome. However, not all packages impact the application
build equally. Some of them are more sensitive than others, and it would
be very helpful for package managers and users to be aware of them. Table~\ref{tab:pkg_imp} shows
the top five packages or parent child pairs (represented as \emph{parent}+\emph{child}, for example autoconf+m4) that are 
most sensitive and is likely to impact the outcome of the build process of the root package. The importance score calculated using Eq~\ref{eq:d_js}. As can be seen
from Table~\ref{tab:package_details}, each root package tens if not hundreds of
dependencies, and the number of package pairs formed by parent and child is on the
order of hundreds. Having the knowledge of which packages have the most influence on the build outcome
can can be of significant help during the build space exploration, especially when
installing on a new platform or looking to upgrade to a newer version.

\subsection{Extracting Package Dependency Constraints}
\begin{figure*}[ht]
\centering
\includegraphics[width=0.98\textwidth]{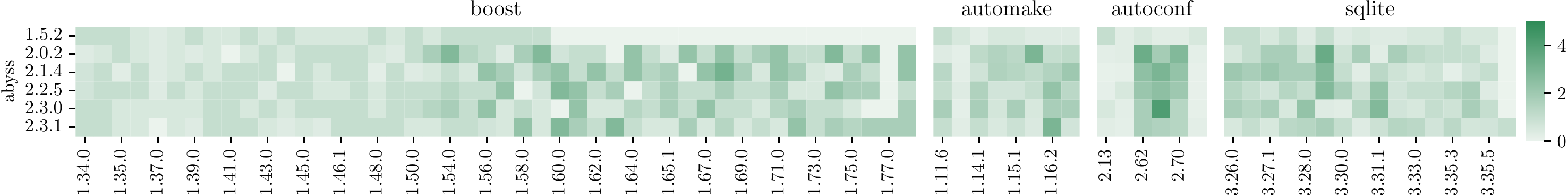}
\caption{Heatmap of Abyss and its dependencies with scores indicating which version pairs are highly likely to build. It shows that
a newer version of Abyss (version greater than 1.5.2) is not likely to build with an older version of Boost (version 1.60.0).}
\label{fig:abyss_score}
\end{figure*}

\begin{figure*}[ht]
\centering
\includegraphics[width=0.98\textwidth]{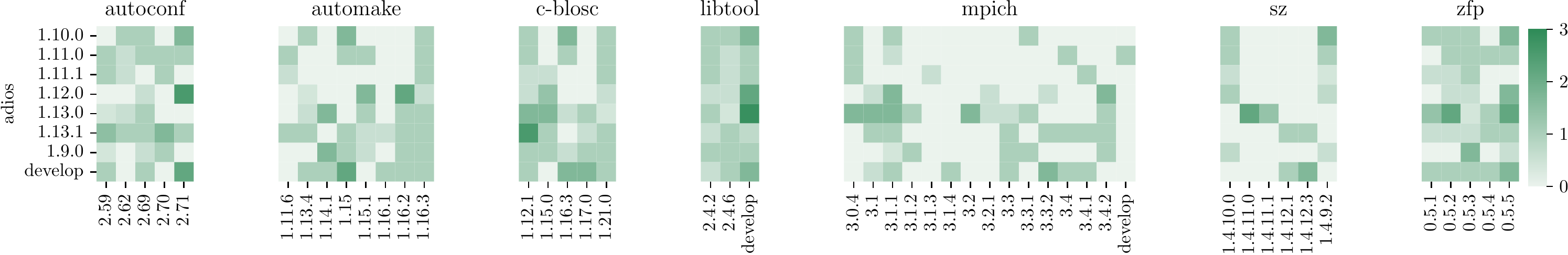}
\caption{Heatmap of Adios and its dependencies with scores indicating which version pairs are highly likely to build.}
\label{fig:adios_score}
\end{figure*}

\begin{figure}[ht]
\centering
\includegraphics[width=0.5\textwidth]{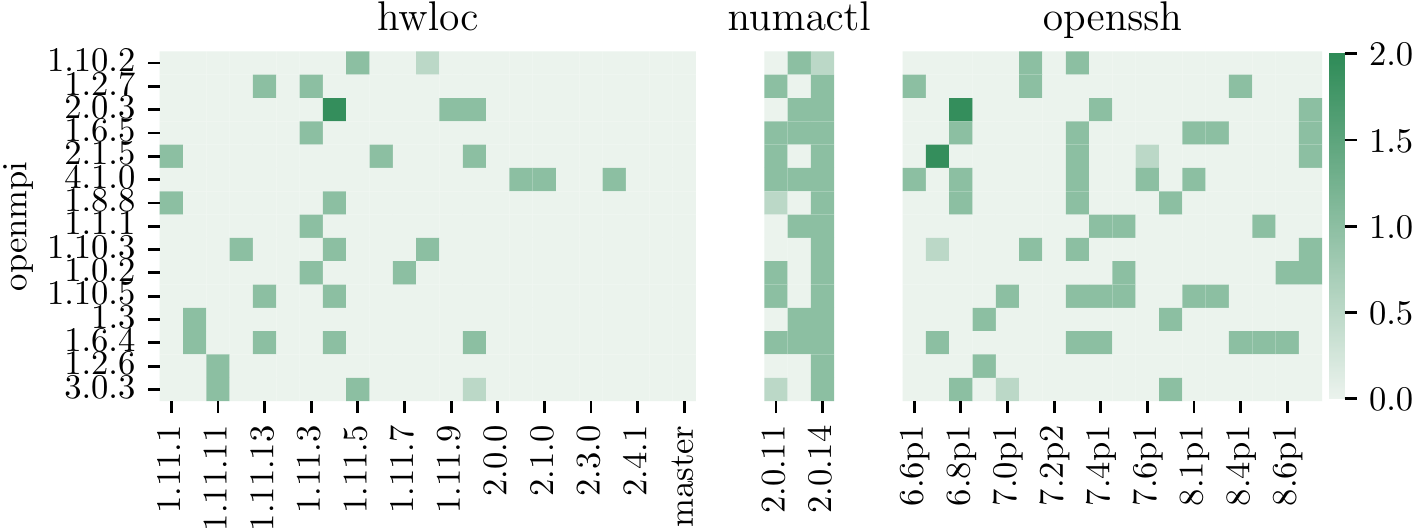}
\caption{Heatmap of OMPI and its dependencies with scores indicating which version pairs are highly likely to build.}
\label{fig:openmpi_score}
\end{figure}

We use all the data from the dataset to build the \emph{Bayesian} model and analyze incompatibility between 
various packages.
We utilize the good and the bad probability densities from Eq~\ref{eq:pxy} 
pertaining to a parent child pair and calculate the $\mathcal{EI}$ from Eq~\ref{eq:eipx}
for the different package version combinations. 
Figures~\ref{fig:abyss_score},~\ref{fig:adios_score},~\ref{fig:openmpi_score} show the heatmap indicating which version pairs are highly likely to build
and which ones are not. A lower score indicates that the pair is not likely to build.
This analysis was done for all the packages, but in the interest of space we show only a few of them here.
Figure~\ref{fig:abyss_score} shows which version pairs are incompatible in the case
of Abyss. One of the insights provided by the figure is that a newer version of Abyss (version greater than 1.5.2) is not likely
to build with an older version of Boost (version 1.60.0). Similarly, we can see from
the figure that an older version of autoconf, such as 2.13, is most likely incompatible 
with versions of Abyss greater than 2.0.2. Moreover, note that the version pair combinations
that are highly likely to build is a much smaller set which indicates that this package
pair is highly sensitive to version changes. As a result, autoconf is among the top $5$
sensitive packages for Abyss as shown in table~\ref{tab:pkg_imp}. Similar analysis can be done
for Adios and OpenMPI based on the figures~\ref{fig:adios_score} and ~\ref{fig:openmpi_score} 
This information can be leveraged by the package manager
to introduce new package dependency constraints so as to avoid configurations that can
result in build failures.

\section{Related Work}

In 1997 the notion of ``software release management''~\cite{van1997software} was introduced for large collections of independent packages. During that period Linux distributions broadly adopted the ~\cite{rpm, apt} package managers. The version selection problem is NP-complete and can be encoded as SAT and Constraint Programming (CP) problems~\cite{dicosmo:edos,mancinelli+:ase06-foss-distros}. 
Since then the idea of customizable solvers provide modular package managers ~\cite{abate2012dependency}. The work focuses on the Common Upgradeability Description Format (CUDF). Commonly, this file format is used by the front-end of package managers to describe upgradeability scenarios to the back end-solvers.
Multiple implementation are proposed to solve the CUDF scenario which employ 
Mixed-Integer Linear Programming, Boolean Optimization, and Answer Set
Programming~\cite{michel+:lococo2010,argelich+:lococo2010,gebser+:2011-aspcud}. 
Such solutions have been adopted by various Linux distributions~\cite{abate2020dependency}.
Today, complete solvers are being broadly adopted by various package managers. For example, PIP recently switched to a new proper solver~\cite{pip-new-resolver}. Dart now uses CDCL SAT solver called {\tt PubGrub}~\cite{weizenbaum:pubgrub18}, and Rust's Cargo~\cite{cargo} package manager is moving towards this approach~\cite{pubgrub-rs}.

Despite the completeness of the modern package manager solvers, the package managers 
rely heavily on a correct pre-selection of version dependencies among packages and 
detection of possible conflicts.
 A popular solution is semantic versioning (\emph{semver})~\cite{preston2013semantic}.
 In semantic version \emph{semver} a version number characterize package compatibility and
 breaking changes.  Semantic versioning heavily relies on developers knowing the rules
of compatibility and versioning their package releases correctly.  Reports and empirical studies~\cite{decan2019package, dietrich2019dependency} indicate a broad adoption of \emph{semver}. However, the rules are complex, and thus not fully understood~\cite{dietrich2016java}, and therefore frequently resulting into \emph{semver} misuse~\cite{ochoa2021breaking}. In the end, developers drop the usage of semantic versioning and use
fixed versioning disregarding all the advantages of flexible versioning schemes. More importantly though, 
as stated in ~\cite{gamblin2021software} if all projects, in the end, follow a fixed versioning approach 
 conflicts will eventually arise that prevent all packages from building.

In~\cite{kula2018developers} the authors study how frequently and how soon package maintainers update their dependencies to include the latest release of dependent libraries. Interestingly, package maintainers rarely update their dependencies, resulting in software systems which contain known vulnerabilities. The results indicate that a heavy reliance on libraries in contemporary projects often result in the formation of complex inter-dependency relationships inside that project. Due to such inter-dependency issues developers are reluctant to update their dependencies. 

There have been multiple approaches that suggest specific dependency versions. The wisdom of the crowd~\cite{mileva2009mining} is a popular approach. Where maintainers depend on the most popular, or highly used,
library versions. Based on such information users and maintainers can more easily select library dependencies and avoid
installation errors. Other tools identify incompatibilities between versions at the binary level~\cite{cossette2012seeking}. In contrast, some propose techniques for 
dependent packages to automatically perform code changes and adopt to conflicts introduced by their dependencies~\cite{xu2019meditor}.
Multiple works~\cite{nguyen2021recommending, ouni2017search, nguyen2020crossrec, nguyen2018mining,he2020diversified,sun2020req2lib} focus on suggesting third party libraries to use in projects. In contrast to our approach, these efforts focus on suggesting new libraries, and thus new dependencies. Our work focuses on a different facet of the problem, how to select which version of an existing third-party library to use in a given configuration. 

These works provide mechanisms to prevent using conflicting libraries and thus avoid 
performing failing installations. In our approach we embrace and accept the existence of failing build configurations.
However, we provide an active learning mechanism through autotuning that learns which configurations tend to fail
and after a reasonable amount of iterations is able to predict high-fidelity built configurations. Moreover, our 
technique is language and system agnostic. In essence it only requires monitoring the final result of the build process. 
\section{Conclusions}
\label{sec:conlusions}
We have presented \tool{}, an active-learning-based framework
to select high-fidelity build configurations using limited number of
samples. We implemented two models, one based on \emph{Wisdom of the crowd}
and another based on \emph{Bayesian} optimization and evaluate their
efficacy in identifying build configurations that are highly likely to build.
For the purpose of our evaluation, we collected a large dataset consisting
of packages from the E4S package ecosystem. We showed
that our models are able to select high fidelity configurations with far
fewer samples in comparison to a random exploration. For example, \tool{} 
selects $3\times$ the number of good configurations in comparison to random sampling 
for several packages including Abyss, Bolt, libnrm, OpenMPI.
Our framework is also able to select all the high-fidelity builds in half the number 
of samples as required by random sampling for packages such as Chai, OpenMPI, py-petsc4py, slepc.
We also leveraged our model
to provide insights about package incompatibilities, which can 
be used by both package managers as well as users to identify problematic version
pairs and avoid including them. Most importantly, \tool{} provides an automatic
way to select high-fidelity build configurations with significantly less samples.
\section{Acknowledgements}
This work was performed under the auspices of the U.S. Department of Energy by Lawrence Livermore National Laboratory under Contract DE-AC52-07NA27344 (LLNL-CONF-831078-DRAFT). Work at LLNL was funded by the Laboratory Directed Research and Development Program under project tracking code 21-SI-005.

\bibliographystyle{plain}
\bibliography{references}
\end{document}